\documentclass[%
prl%
 ,secnumarabic%
,tightenlines%
,amssymb, amsmath, nobibnotes, nofootinbib,showpacs, aps]{revtex4}
\usepackage{docs}%
\usepackage{bm}%
\usepackage{epsfig}%
\usepackage{graphicx}%
\expandafter\ifx\csname package@font\endcsname\relax\else
 \expandafter\expandafter
 \expandafter\usepackage
 \expandafter\expandafter
 \expandafter{\csname package@font\endcsname}%
\fi

\begin{document}

\title{GSI anomaly and spin-rotation coupling}%

\author{Gaetano Lambiase$^{a,b,c}$, Giorgio Papini$^{c,d,e}$, Gaetano Scarpetta$^{a,b,c}$}
\affiliation{$^a$Dipartimento di Fisica "E.R. Caianiello"
 Universit\'a di Salerno, 84081 Baronissi (Sa), Italy.}
  \affiliation{$^b$INFN, Sezione di Napoli Italy.}
  \affiliation{$^c$International Institute for Advanced Scientific Studies, 89019 Vietri sul Mare (SA), Italy.}
  \affiliation{$^d$Department of Physics, University of Regina, Regina, SK, S4S 0A2, Canada.}
  \affiliation{$^e$Prairie Particle Physics Institute, Regina, SK, S4S 0A2, Canada}
\def\be{\begin{equation}}
\def\ee{\end{equation}}
\def\al{\alpha}
\def\bea{\begin{eqnarray}}
\def\eea{\end{eqnarray}}

\begin{abstract}

We propose a model in which a recently reported modulation in the decay of the hydrogenlike ions ${}^{140}$Pr$^{\, 58 +}$,
${}^{142}$Pm$^{\, 60 +}$ and ${}^{122}$I$^{\, 52 +}$ arises from the coupling of rotation to the
spin of electron and nucleus.
The model shows that the spin-spin coupling of electron and nucleus does not contribute to the
modulation and predicts that the anomaly cannot be observed if
the motion of the ions is rectilinear, or if the ions are stopped in a
target. It also supports the notion that the modulation frequency is proportional to the inverse of the atomic mass and that
no modulation is expected for the $\beta^+$-decay.
\end{abstract}


\maketitle

\section{Introduction}

Experiments carried out at the storage ring ESR of GSI in Darmstadt \cite{GSI,GSIb,geissel} reveal an oscillation in the
orbital electron capture and subsequent decay of hydrogenlike
${}^{140}$Pr$^{\, 58 +}$, ${}^{142}$Pm$^{\, 60 +}$ and $ {}^{122}$I$^{\, 52 +}$. The
modulation has a period of $7.069(8)\,$s, $7.10(22)\,$s and $6.1\,$s
respectively in the laboratory frame and is superimposed on the
expected exponential decay. The "zero hypothesis" of a pure experimental decay has been excluded
at the $99\%$ C.L. and periodic instabilities in the storage ring and detection apparatus
also seem improbable causes of the modulation. The effect has
been extensively studied in literature \cite{mixing,giuntiPLB,merleIOP,merlePRC}.

We show, in the model proposed below, that a modulation arises in
the probability that the system, initially in a superposition of
hyperfine states ($F=3/2$ and $F=1/2$), finds itself again in such
a superposition of hyperfine states after injection into the
storage ring. The modulation has its origin in the spin-dependent
part of the Thomas precession, and is compatible with the observed
ESR modulation. The EC decay occurs for states with spin $F=1/2$
because decay from the spin $3/2$ state is forbidden by the
conservation of the $F$ quantum number \cite{GSIb}. We stress that
the present paper differs in essential ways from \cite{lambiase08}
because it takes into account all the relevant features of the GSI
experiment, such as bound states kinematics, dragging effects,
Thomas precessions of nucleus and electron and QED and derives the
probability of the observed modulation from the time evolution of
nucleus plus electron once this system is injected in the storage
ring.

The full Hamiltonian that describes the behavior
of nucleus and bound electron in the external field ${\bf B}$ of the ring is
$H=H_0+H_1$, where $H_0$ contains all the usual standard terms (Coulomb potential, spin-orbit coupling,
etc.), and $H_{1}$ is (in units $\hbar=c=1$)
 \begin{equation}\label{H1}
H_1=-{\cal A}\,  {\bf s} \cdot {\bf I}- {\bf s}\cdot {\bm
\Omega}_e-{\bf I}\cdot {\bm \Omega}_n\,,
 \end{equation}
where ${\cal A} \simeq Z^3 \displaystyle{\frac{4\alpha^4 g_n}{3}}\sim N\times 10^{14}\mbox{Hz}\,, N\sim {\cal O}(1)$
is the strength of spin-spin coupling, while
  \begin{eqnarray}
{\bm \Omega}_e  \equiv {\bm\omega}_{g_e}+{\bm\omega}_{Th}^{(e)} -{\bm \omega}_c^{(e)}\,, \label{Omega-e-def} \\
{\bm \Omega}_n \equiv {\bm \omega}_{g_n}+{\bm \omega}_{Th}^{(n)}-{\bm \omega}_c^{(n)} \label{Omega-n-def}\,,
  \end{eqnarray}
represent the precession of the electron spin and the usual spin precession of the nucleus in its motion in a storage
ring that is assumed circular for simplicity. In (\ref{Omega-e-def}) and (\ref{Omega-n-def}), ${\bm \omega}_{g_{e,n}}$ are the electron and nucleus
spin precession frequencies due to the respective magnetic moments $g_{e,n}$ and
${\bm \omega}_c^{(e,n)}$ are the angular cyclotron frequencies.
The explicit expressions of all these quantities are given below.
We refer (\ref{H1}) to a frame rotating about the $x_3$-axis
in the clockwise direction of the ions, with the $x_2$-axis
tangent to the ion orbit in the direction of its momentum
and write ${\bf B}=B{\bf \hat u}_3$, where $B=1.197$T is the GSI value
(we are assuming that this is  the average value over the circumference).

The Thomas precession ${\bm \omega}_{Th}^{(e,n)}$ is related to
the standard spin-rotation coupling that can be derived, for the
electron, from the spin connection coefficients of the Dirac
equation in a rotating frame \cite{jackson,pap}. In our
derivation, we neglect any stray electric fields and electric
fields needed to stabilize the nucleus orbits, as well as all
other effects which could affect the Thomas precession
\cite{silenko}.

We indicate by ${\bm \beta}$ and ${\bm \beta}_n$ the velocities of electron and nucleus relative to the lab frame.
Using the composition of
velocities, the Lorentz factor $\gamma=1/\sqrt{1-{\bm\beta}^2}$ of the electron can be written in the form $\gamma=\gamma_{n}\gamma_{e|n}(1+\Pi)$, where $\Pi={\bm
\beta_n}\cdot {\bm \beta_{e|n}}=\beta_{n}\beta_{e|n}\cos \theta$, ${\bm \beta}_{e|n}$ is the velocity of the
electron relative to the nucleus, $\gamma_{n}=1/\sqrt{1-{\bm\beta}_n^2}$,
and $\gamma_{e|n}=1/\sqrt{1-{\bm\beta}_{e|n}^2}$. The explicit expression of ${\bm \beta}$ is also useful
 \be\label{beta}
 {\bm \beta}= \frac{1}{\gamma_n(1+\Pi)}
 \left[{\bm\beta}_{e|n}+\frac{\gamma_n^2\Pi}{\gamma_n+1}\, {\bm \beta}_n +\gamma_n {\bm \beta}_n\right]\,.
 \ee
The Thomas precession of the electron in the lab frame is given by
${\bm\omega}_{Th}=-\displaystyle{\frac{\gamma^2}{\gamma+1}\frac{d{\bm\beta}}{dt}\wedge
{\bm\beta}}$. The field $\bf B$ in the lab frame (where $\bf E
=0$) is transformed to the nucleus rest frame and gives ${\bf E}'=
\gamma_n {\bm \beta}_n\wedge {\bf B}$ and ${\bf B}'=\gamma_n {\bf
B}$ on account of ${\bm \beta}_n \cdot {\bf B}=0$. The equations
of motion are $\displaystyle{\frac{d{\bm\beta}_{e|n}}{dt_n}}=
 \displaystyle{\frac{{\bf f}_{e|n}}{\gamma_{e|n} m}}-
\displaystyle{ \frac{{\bm \beta}_{e|n}}{ \gamma_{e|n} } \frac{{\bm \beta}_{e|n}\cdot {\bf f}_{e|n}}{m}}$ for the electron with respect to the nucleus and $\displaystyle{\frac{d{\bm\beta}_{n}}{dt}}=\displaystyle{\frac{Q}{M\gamma_n}}\,{\bm \beta}_n\wedge {\bf B}$ for the nucleus with respect to the lab frame. Here ${\bf f}_{e|n}=-e ({\bf E}'+{\bm \beta}_{e|n}\wedge {\bf B}')=-e\gamma_n ({\bm \beta}_{e|n}+{\bm \beta}_{n})\wedge {\bf B}$.
Using $dt=\gamma_n (1+\Pi) dt_n$, taking ${\bm \beta}_{e|n}\cdot {\bf B}=0$, ${\bf E}_{e|n}\wedge {\bf \beta}_{e|n}=0$ and ${\bf E}_{e|n}\wedge {\bf \beta}_{n}=0$ (averaged over the decay time of the ion in the storage ring), we find $\displaystyle{\frac{d{\bm\beta}_{e|n}}{dt}}=-\displaystyle{\frac{e}{m}\frac{1}{\gamma_{e|n}(1+\Pi)}
 [{\bm \beta}_{e|n}+{\bm \beta}_{n}]\wedge {\bf B}}$.
Neglecting spin-orbit coupling\footnote{The Coulomb interaction also contributes to the Thomas precession. It generates the spin-orbit coupling term in the Hamiltonian of the electron $H_C\sim \frac{g_e-1}{2m_e r}(dV/dr) {\bf s}\cdot {\bf L}$. However ${\bf s}\cdot {\bf L}=\frac{1}{2}[j(j+1)-l(l+1)-s(s+1)]$ vanishes when $j=s$ for $l=0$ and $j=l\pm 1/2$  for $l\neq 0$, therefore $H_C=0$ in the ground state.
Moreover, the effect of ${\bf A}$ in ${\bm \pi}=\frac{1}{m}({\bf p}-e{\bf A})$
is negligible in the present context.},
${\bm\omega}_{Th}$ can be written as
 \begin{equation}\label{thomas-fin}
 {\bm\omega}_{Th}
 =  \frac{e{\bf B}}{m}\frac{1}{\gamma_{e|n}\gamma_n} I_e
 - \frac{Q{\bf B}}{M} \frac{1}{\gamma_n} I_Q\,,
   \end{equation}
where
 \[
 I_e\equiv \frac{\displaystyle{(\gamma_{e|n}\gamma_n)^2\left({\bm \beta}_n^2+\frac{{\bm \beta}_{e|n}^2}{\gamma_n}+2\Pi+\frac{\gamma_n \Pi^2}{\gamma_n+1}-Y\right)}}{(1+\Pi)[\gamma_{e|n}\gamma_n(1+\Pi)+1]}\,,
 \]
 \[
 I_Q\equiv \frac{(\gamma_{e|n}\gamma_n)^2
 \displaystyle{
 \left[{\bm \beta}_n^2\left(1+\frac{\gamma_n\Pi}{\gamma_n+1}\right)^2-X\right]}}{\gamma_{e|n}\gamma_n(1+\Pi)+1}\,,
 \]
 \[
 Y\equiv \frac{{\bm \beta}_{e|n}^2[\gamma_n(2-\cos^2\theta)-\sin^2\theta]}{3\gamma_n^2},
 X\equiv \frac{{\bm \beta}_{e|n}^2{\bm \beta}_{n}^2\sin^2\theta}{3(\gamma_n+1)}\,.
 \]
The coupling of the electron magnetic moment ${\bm \mu}_e=-\displaystyle{\frac{g_e}{2}\frac{e}{m_e}}{\bf s}$ with the magnetic field is described by
 \[
 H_{g_e}=\frac{1}{\gamma}{\bm \mu}_e\cdot {\bf B}''={\bm\mu}_e\cdot \left[{\bf B}-\frac{\gamma}{\gamma+1}{\bm \beta}({\bm \beta}\cdot{\bf B})\right]\,.
 \]
Keeping only the quadratic term in ${\bm \beta}_{e|n}$ and using
${\bm \beta}({\bm \beta}\cdot{\bf B})=\displaystyle{\frac{({\bm \beta}_{e|n}\cdot {\hat {\bf u}_3})^2}{\gamma_n^2 (1+\Pi)^2}\, {\bf B}}$,
we obtain
 \begin{equation}\label{Hge}
 H_{g_e}=\Upsilon \, {\bm\mu}_e\cdot {\bf B}\,,
 \quad \Upsilon \equiv 1-\frac{\gamma_{e|n}^2({\bm \beta}_{e|n}\cdot {\hat {\bf u}_3})^2}{\gamma(\gamma+1)}\,,
 \end{equation}
and from it $d{\bf s}/dt =i[H_{g_e}, {\bf s}]={\bm \omega}_{g_e}\wedge {\bf s}={\Upsilon}{\bm \mu}_e\wedge {\bf B}$ which yields
 \begin{equation}\label{omegage}
{\bm \omega}_{g_e}=-\displaystyle{\frac{g_e e}{2m_e}\,{\Upsilon}\, {\bf B}}\,.
 \end{equation}
In order to refer the spin precession to the particle orbit, the effective cyclotron
frequency ${\bm \omega}_c^{(e)}$ must now be subtracted.
Its value is obtained by computing the
instantaneous acceleration $d{\bm \beta}/dt=\omega_e^{(c)}\beta {\hat{\bf u}}_1$.
Omitting terms like ${\bf a}_{e|n} = \frac{q}{m_e} {\bf E}_{e|n}$, ${\bm\beta}_{e|n} \wedge {\bf B}$ and
$[{\bf B}\cdot ({\bm \beta}_{e|n}\wedge {\bm \beta}_n)]{\bm \beta}_n$ that vanish when averaged, as already pointed out,
we find
 \begin{equation}\label{omegace}
 {\bm \omega}_c^{(e)}=\left[-\frac{eB}{m_e}\frac{\beta_n}{\beta}\frac{1-({\bm \beta}_{e|n}\cdot {\hat {\bf u}}_1)^2}{\gamma_{e|n}\gamma_n(1+\Pi)^2}
  + \frac{QB}{M} \frac{\beta_n}{\gamma_n \beta} \Xi_n \right]{\bf u}_3\,,
 \end{equation}
where $\Xi_n\equiv \displaystyle{\frac{1}{\gamma_n+1}\left(1+\frac{\gamma_n \Pi}{\gamma_n+1}\right)}$ and
 \[
 \beta =\sqrt{\frac{{\bm \beta}_n^2+\frac{{\bm \beta}_{e|n}^2}{\gamma_n^2}+2\Pi +\Pi^2}{(1+\Pi)^2}} \,.
 \]
From (\ref{Omega-e-def}),(\ref{thomas-fin}) and (\ref{omegace}) we obtain
 \begin{equation}\label{Omega_e}
  {\bm \Omega}_e  = -\frac{e {\bf B}}{m_e}\left(\frac{g_e}{2} {\Upsilon} -\frac{I_e}{\gamma_{e|n}\gamma_n}-U\right)
  -\frac{Q{\bf B}}{M}\frac{I_Q + V}{\gamma_n}\,,
 \end{equation}
where $\Upsilon$ is defined in (\ref{Hge}) and
 \begin{equation}\label{Upsilon-U-V}
 U\equiv \frac{1-({\bm \beta}_{e|n}\cdot {\hat {\bf u}_1})^2}{\gamma_{e|n}\gamma_n(1+\Pi)^2}\frac{\beta_n}{\beta}\,,
 V \equiv \frac{\beta_n}{\beta(1+\Pi)}\left(1+\frac{\gamma_n \Pi}{\gamma_n+1}\right)\,.
 \end{equation}
Notice that the standard result $\Omega_e = -e {\bf B}a_e/m_e$,
where $a_e=(|g_e|-2)/2$ is the electron magnetic moment anomaly, is recovered in the limit $Q=0$.

The calculation of $g_e$-factors, based on bound state (BS) QED, can be carried out with accuracy even though, in our case, the
expansion parameter is $Z\alpha\simeq 0.4$. The BS-QED calculation gives \cite{vogel,blundell}
\begin{equation}\label{gb}
g_e^b =2\left[\frac{1+2\sqrt{1-(\alpha
Z)^2}}{3}+\frac{\alpha}{\pi}C^{(2)}(\alpha Z)\right]\,,
\end{equation}
where $C^{(2)}(\alpha Z)\simeq \frac{1}{2}+\frac{1}{12}(\alpha Z)^2
+\frac{7}{2}(\alpha Z)^4$.
From (\ref{gb}) we obtain the values  $a_e \simeq
-0.065122$, $a_e \simeq -0.0682112$ and $a_e=0.0505352$ for
${}^{140}$Pr$^{\, 58 +}$, ${}^{142}$Pm$^{\, 60 +}$ and ${}^{122}$I${}^{\, 52 +}$ respectively.
The addition of more expansion terms
\cite{CODATA} does not change these results appreciably.

Consider now the nucleus with spin ${\bf I}$. The terms of (\ref{Omega-n-def}) are ${\bm \omega}_{g_n}=\displaystyle{g_{n}\mu_{N}} {\bf B}$, where $\mu_{N}=\displaystyle{\frac{|e|}{2 m_{p}}}$, ${\bm \omega}_{Th}^{(n)}=-\displaystyle{\frac{\gamma_n-1}{\gamma_n}\frac{Q{\bf B}}{M}}$, ${\bm \omega}_c^{(n)}=\displaystyle{\frac{Q{\bf B}}{M\gamma_n}}$ and give
 \begin{equation}\label{Omega_n}
 {\bm \Omega}_n = \frac{Q{\bf B}}{M}\left(\frac{g_n}{2}\frac{A}{Z}-1\right)\,.
 \end{equation}

\section{Probability and modulation}

Let $|I, m_I\rangle_I$ and $|s,
m_s\rangle_s$ be the eigenstates of the operators ${\bf \hat I}$ and
${\bf \hat s}$. The total angular momentum operator is ${\bf \hat
F}={\bf \hat s}+{\bf \hat I}$. The angular momentum $F$ assumes the values $F=
3/2, 1/2$, $m_{F=3/2}=\pm 3/2, \pm 1/2$, and $m_{F=1/2}=\pm 1/2$
because $I=1$, $m_I=\pm 1, 0$ and $s=1/2$, $m_s=\pm 1/2$. By
making use of the raising and lowering operators ${\bf \hat
F}_\pm={\bf \hat I}_\pm+{\bf \hat s}_\pm$, we construct the
normalized and orthogonal states
 \begin{eqnarray}
 \phi_1&\equiv& \Big|\frac{3}{2},\frac{3}{2}\rangle_F=|1,1\rangle_I
 \Big|\frac{1}{2},\frac{1}{2}\rangle_s \nonumber
   \end{eqnarray}
 \begin{eqnarray}
 \phi_2 & \equiv & \Big|\frac{3}{2},\frac{1}{2}\rangle_F=\sqrt{\frac{2}{3}}|1,0\rangle_I
 \Big|\frac{1}{2},\frac{1}{2}\rangle_s +\sqrt{\frac{1}{3}}|1,1\rangle_I
 \Big|\frac{1}{2},-\frac{1}{2}\rangle_s \nonumber
  \end{eqnarray}
 \begin{eqnarray}
  \phi_3 & \equiv &  \Big|\frac{3}{2},-\frac{1}{2}\rangle_F=  \nonumber  \\
    & = & \sqrt{\frac{2}{3}}|1,0\rangle_I
 \Big|\frac{1}{2},-\frac{1}{2}\rangle_s +\sqrt{\frac{1}{3}}|1,-1\rangle_I
 \Big|\frac{1}{2},\frac{1}{2}\rangle_s \nonumber
   \end{eqnarray} 
   \begin{eqnarray}
 \phi_4 & \equiv & \Big|\frac{3}{2},-\frac{3}{2}\rangle_F=|1,-1\rangle_I
 \Big|\frac{1}{2},-\frac{1}{2}\rangle_s \nonumber
   \end{eqnarray} 
 \begin{eqnarray}
 \phi_5 & \equiv & \Big|\frac{1}{2},\frac{1}{2}\rangle_F=\sqrt{\frac{1}{3}}|1,0\rangle_I
 \Big|\frac{1}{2},\frac{1}{2}\rangle_s - \sqrt{\frac{2}{3}}|1,1\rangle_I
 \Big|\frac{1}{2},-\frac{1}{2}\rangle_s \nonumber \\
 \phi_6 & \equiv & \Big|\frac{1}{2},-\frac{1}{2}\rangle_F= \nonumber \\
  & = & -\sqrt{\frac{2}{3}}|1,-1\rangle_I
 \Big|\frac{1}{2},\frac{1}{2}\rangle_s + \sqrt{\frac{1}{3}}|1,0\rangle_I
 \Big|\frac{1}{2},-\frac{1}{2}\rangle_s \nonumber\,.
  \end{eqnarray}
The ($6\times 6$) matrix with elements $\langle \phi_i | {\hat
H}_1 | \phi_j \rangle$ 
has the eigenvalues
 \begin{eqnarray}
 \lambda_{1,4}&=& -\frac{\cal A}{2} \pm \left(\frac{\Omega_e}{2}+\Omega_n\right)\,, \nonumber \\
 \lambda_{2,3} &=& \frac{\cal A}{4} \mp \frac{\Omega_n}{2} - \frac{\sqrt{\Delta_\pm}}{4}\,, \,\,
 \lambda_{5,6} = \frac{\cal A}{4} \mp \frac{\Omega_n}{2}+\frac{\sqrt{\Delta_\pm}}{4}\,, \nonumber
 \end{eqnarray}
where
 \[
\Delta_\pm = 9{\cal A}^2\pm 4 {\cal A}\Omega_e \mp 4{\cal A}\Omega_n +4(\Omega_e-\Omega_n)^2\,,
 \]
and the corresponding eigenstates $|i\rangle$ ($i=1, \ldots, 6$)
\begin{eqnarray}
 |1,4\rangle &=& \phi_{1, 4}\,, \quad
 |2, 5\rangle = \frac{B_\pm}{\sqrt{1+B_\pm^2}}\phi_2+ \frac{1}{\sqrt{1+B_\pm^2}}\phi_5 \,, \nonumber \\
 |3, 6\rangle &=& -\frac{A_\pm}{\sqrt{1+A_\pm^2}}\phi_3+ \frac{1}{\sqrt{1+A_\pm^2}}\phi_6 \,, \nonumber
\end{eqnarray}
where
 \[
 A_{\pm} = \displaystyle{ \frac{9{\cal A}-2\Omega_e+2\Omega_n\pm 3\sqrt{\Delta_-}}{4\sqrt{2}\, (\Omega_e-\Omega_n)} }\,,
 \]
 and
 \[
 B_{\pm} = \displaystyle{ \frac{9{\cal A}+2\Omega_e-2\Omega_n\pm 3\sqrt{\Delta_+}}{4\sqrt{2}\, (\Omega_e-\Omega_n)} }\,.
 \]
In the limit ${\cal A}\gg \Omega_{e, n}$ we obtain
 \begin{equation}\label{states}
| i\rangle \simeq |\phi_i \rangle\,,
 \end{equation}
and
 \begin{equation}\label{lambda1-3}
 \lambda_1-\lambda_2=\lambda_2-\lambda_3=\frac{\lambda_1-\lambda_3}{2}=-\frac{\Omega_e+2\Omega_n}{3}\,,
 \quad \lambda_1-\lambda_4=-\Omega_e+2\Omega_n\,,
 \end{equation}
 \begin{equation}\label{lambda5-6}
  \lambda_5-\lambda_6=\frac{\Omega_e-4\Omega_n}{3}\,.
 \end{equation}
Notice that in these expressions the ${\cal A}$-terms coming from the spin-spin coupling {\it cancel out}.

\subsection{Modulation induced by quantum beats}

These results must be now applied to the GSI experiment. Since the heavy nucleus decays via EC, only the states
with $F=1/2$ are relevant. For simplicity we confine ourselves to the Hilbert subspace spanned by the states $\{ |5\rangle, |6\rangle \}$.
Here we follow \cite{giuntiPLB} (see also \cite{merleIOP,merlePRC}).
The decay processes involved in the GSI experiment
${}^{140}\text{Pr}^{\, 58 +}\to {}^{140}\text{Ce}^{\, 58 +} + \nu_e$,
${}^{142}\text{Pm}^{\, 60 +} \to {}^{140}\text{Nd}^{\, 58 +} + \nu_e$, and
${}^{122}\text{I}^{\, 52 +} \to {}^{122}\text{Te}^{\, 52 +} + \nu_e$, can be schematically represented as
\begin{equation}\label{GSIprocess}
    {\mathbb{I}} \quad \to \quad {\mathbb{F}}+\nu_e\,,
\end{equation}
with obvious meaning of the symbols. At the initial instant $t=0$ (before injection into the ESR) the
system nucleus-electron
is produced in a superposition  of the states $\{ |5\rangle, |6\rangle \}$,
 \[
|{\mathbb{I}}(0) \rangle=\sum_{a=5}^6 c_a |a \rangle= c_5 |5 \rangle +c_6 |6 \rangle \,.
 \]
with $|c_5|^2+|c_6|^2=1$. 
If one assumes, for simplicity, that the two states with energies $\lambda_5$ and $\lambda_6$
decay with the same rate $\Gamma$, at the time $t$ the system evolves to the state
 \[
|{\mathbb{I}}(t)\rangle=e^{-\Gamma t/2}\left( c_5 e^{-i \lambda_5 t} |5 \rangle +c_6 e^{-i \lambda_6 t} |6 \rangle \right)\,.
 \]
The probability of EC at time $t$ reads
 \begin{equation}
 P_{EC}(t) =  e^{-\Gamma t} |\langle \nu_e, {\mathbb{F}}|S|{\mathbb{I}}(t)\rangle|^2 =
 e^{-\Gamma t} {\bar P}_{EC}\left[1+a_{56}\cos \left(\omega_{56} t + \varsigma\right)\right]\,, \label{prob}
 \end{equation}
where (see Eq. (\ref{lambda5-6}))
 \begin{equation}\label{omega-F}
\omega_{56} = |\lambda_5-\lambda_6| = \frac{\Omega_e-4\Omega_n}{3}\,,
 \end{equation}
${\bar P}_{EC}=|\langle \nu_e, {\mathbb{F}}|S|5\rangle|^2=|\langle \nu_e, {\mathbb{F}}|S|6\rangle|^2$,
$a_{56}=2|c_5| |c_6|$, and finally $S$ is the interaction operator\footnote{If we consider
the Hilbert space spanned by the states (\ref{states}),
$\{ |1\rangle,\ldots, |6\rangle\}$, then the probability (\ref{prob}) assumes the form
$P_{EC}(t) \sim [1+\sum_{i<j} a_{ij}\cos\omega_{ij}t]$,
with $\omega_{ij}=|\lambda_j-\lambda_j|$and $i, j=1, 5$. It contains 5 terms of which
only one (that due to (\ref{lambda1-3}) and (\ref{lambda5-6})) contributes to the probability,
while the others vanish because of EW selection rules.
Assuming, therefore, that the states are equiprobable, the magnitudes take the value ${\tilde a}_{ij}
=\frac{1}{5}\simeq 0.2$, with $(i,j) = \{(1,2), (1,3), (1,4), (2,3),
(5,6)\}$. The values obtained in the GSI experiments are:
$a$(Pr) = 0.18(3), $a$(Pm) = 0.23(4), $a$(I) = 0.22(2) \cite{kienle}.}.
The phase $\varsigma$ comes form possible phase differences of the amplitude $c_1$ and $c_2$ and of
$|\langle \nu_e, {\mathbb{F}}|S|5\rangle|$ and $|\langle \nu_e, {\mathbb{F}}|S|6\rangle|$.
As (\ref{prob}) and (\ref{omega-F}) show, the modulation of the decay probability
{\it does not} depend on ${\cal A}$.


\subsection{Estimate of $\gamma_{e|n}$}

\begin{figure}[tr]
\resizebox{8cm}{!}{\includegraphics{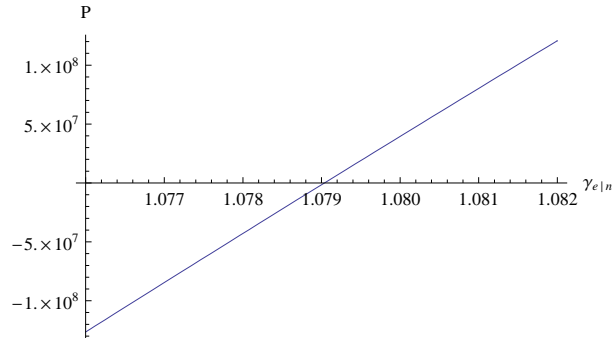}} \caption{P is
defined by
 P $ \equiv \frac{|e|{\bf B}}{3m_p}\Big[\frac{m_p}{m_e}  \Big({\tilde \Upsilon}(a_e+1)-\frac{I_e}{\gamma_{e|n}\gamma_n}-{\tilde U}\Big)+$
  $\frac{I_Q+{\tilde V}}{\gamma_n}\frac{Z}{A}
 +4\frac{Z}{A}\Big(\frac{g_n}{2}\frac{A}{Z}-1\Big)\Big]  - 2 \pi\, 0.14$ Hz (\ref{Omega-f}),
with ${\tilde \Upsilon}$, ${\tilde U}$ and ${\tilde V}$ given by
(\ref{Upsilon-U-V}). The value of the {\it unknown} $\gamma_{e|n}$ is obtained from $P=0$ by using the experimental data and corresponds to the central value of the range $6.98 s \lesssim T \lesssim 7.06 s$
The plot refers to Pr. Similar plots can
be obtained for Pm and I.}\label{Pr}
\end{figure}

We now compare the frequencies
$\omega_{56}/2\pi$, given by (\ref{omega-F}), with the
experimental values $\sim 0.14\,$ Hz found for ${}^{140}$Pr$^{\,
58 +}$ and ${}^{142}$Pm$^{\, 60 +}$ and $ \sim 0.16\,$Hz for
${}^{122}$I${}^{52 +}$ and consider first the case
$\omega_{56}=\lambda_5-\lambda_6$. We find
 \begin{equation}\label{Omega-f}
 \frac{|e| B}{3m_p}\left[\frac{m_p}{m_e}  \left({\bar \Upsilon}(a_e+1)-\frac{{\bar I}_e}{\gamma_{e|n}\gamma_n}-{\bar U}\right)+
 \frac{I_Q+{\bar V}}{\gamma_n}\frac{Z}{A}
 +4\frac{Z}{A}\left(\frac{g_n}{2}\frac{A}{Z}-1\right)\right] = 2 \pi\, 0.14 \,\mbox{Hz}\,,
 \end{equation}
where a bar on top means average values. These are computed by first expanding the quantities $I_{e, Q}$,
$\Upsilon$, $U$ and $V$ in terms of $\Pi<1$ and then averaging over the angle by means of $\langle \cos^n \theta \rangle =\frac{1+(-1)^n}{2(n+1)}$.
Using $({\bm \beta}_{e|n}\cdot {\hat {\bf u}}_i)^2=\frac{1}{3}{\bm \beta}_{e|n}^2$, $i= 1, 2, 3$,
$\gamma_n^{\text{(Pr, Pm, I)}}=1.43$, $g_{n}^{\text{(Pr,Pm)}}=2.5$, $g_{n}^{\text{(I)}}=0.94$, up to ${\cal O}(\Pi^6)$ we obtain from (\ref{Omega-f}) the numerical solutions
(see Fig. \ref{Pr})
 \begin{equation}\label{gamma}
\gamma^{(\text{Pr})}_{e|n}\sim 1.07904 \,, \quad  \gamma^{(\text{Pm})}_{e|n} \sim 1.08435 \,,
 \quad  \gamma^{(\text{I})}_{e|n}\sim 1.05902 \,,
 \end{equation}
which must be compared with the Lorentz factors of the bound electron in the Bohr model
$\gamma^{(\text{Pr})}_{e|n}\sim 1.0970$, $\gamma^{(\text{Pm})}_{e|n}\sim 1.1040$, and
$\gamma^{(\text{I})}_{e|n}\sim 1.0776$.
The values (\ref{gamma}) imply that the binding energies $E=T+E_p=-m [1-\gamma_{e|n}+(\alpha Z)^2]$, where $T$ ad $E_p$ are
kinetic and potential energies of the bound electron, are given by
 \[
 E^{(\text{Pr})} \sim -54.4 \text{keV}, \quad E^{(\text{Pm})} \sim -58.2 \text{keV}, \quad E^{(\text{I})} \sim -46.3 \text{keV},
 \]
in agreement with the values
 \begin{equation}\label{bindexp}
 E^{(\text{Pr})}_{R} = -49.5\text{keV}, \quad E^{(\text{Pm})}_{R} = -53.1\text{keV}, \quad E^{(\text{I})}_{R} = -39.6\text{keV},
 \end{equation}
derived from the relativistic equation \cite{zuber}
 \begin{equation}\label{dirac}
E_{R}=-\displaystyle{\frac{RZ^2}{n^2}\left[1+\frac{(\alpha Z)^2}{n}\left(1-\frac{3}{4n}\right)\right]}\,,
 \end{equation}
where $R=13.6057$eV and $n=1$.

\section{Conclusions}

In this paper we explain the GSI anomaly by
means of a semiclassical model based on the Thomas precession of
spins. The model has the following consequences: {\it 1)} It
avoids all criticisms raised in \cite{faber} and in \cite{fastermann} because the
Hamiltonians are essentially different. {\it 2)} There is no
modulation in the $\beta^+$-decay branch \cite{kienle}. This
because the Thomas precession, when computed only for a decaying
charged nucleus gives rise to a frequency $\Omega_n \sim eB/m_p
\sim 10^7$ Hz and the probability $\sim e^{- \Gamma t} [1+a
\cos(\Omega_n t)]$. The high frequency modulation term averages
out to zero, and the probability obeys the standard exponential
decay. {\it 3)} The GSI oscillations disappear when $B=0$. {\it
4)} The model is consistent with experiments on EC decays of
neutral atoms in solid environments that have shown no
oscillations/modulations \cite{vetter}. {\it 5)} The model
predicts that $\omega_{56}\sim A^{-1}$ if the
three terms on the l.h.s. of (\ref{Omega-f}) are of the same order
of magnitude (notice however that the $A$-terms only affect the
third digit of $\gamma_{e|n}$ and are not, therefore, a real
discriminating feature of our model). {\it 6)} The model is
consistent with the absence of a periodic transfer from active
$(F=1/2)$ to sterile $(F=3/2)$ states. 

\end{document}